# Active modulation of surfactant driven flow instabilities by swarming bacteria

Harshitha S. Kotian[1][2], Shalini Harkar[2], Shubham Joge[3], Ayushi Mishra[3], Amith Z. Abdulla[1], Varsha Singh[3], Manoj M. Varma[1][2]*

[1] Centre for Nano Science and Engineering, Indian Institute of Science, Bangalore, India
[2] Robert Bosch Centre for Cyber Physical Systems, Indian Institute of Science, Bangalore, India
[3] Molecular Reproduction, Development and Genetics, Indian Institute of Science, Bangalore, India
*mvarma@iisc.ac.in

Models based on surfactant driven instabilities have been employed to describe pattern formation by swarming bacteria. However, by definition, such models cannot account for the effect of bacterial sensing and decision making. Here we present a more complete model for bacterial pattern formation which accounts for these effects by coupling active bacterial motility to the passive fluid dynamics. We experimentally identify behaviours which cannot be captured by previous models based on passive population dispersal and show that a more accurate description is provided by our model. It is seen that the coupling of bacterial motility to the fluid dynamics significantly alters the phase space of surfactant driven pattern formation. We also show that our formalism is applicable across bacterial species.

Surfactant laden drop spreading on a pre-existing film is a well-studied system [1–3] which is characterised by thinning of the film with a thickened corona at the propagating front of the drop. Velocity of the spreading drop is dependent on the local height and surface tension gradient. Any disturbance which modifies the local surfactant gradients at the edge can destabilise the front and lead to the formation of finger-like patterns at the interface [1]. The fingering so formed resemble the viscous fingering in a Hele-Shaw cell [4], though the physics involved is quite different.

Studying surfactant driven films and their stability is crucial to understand many biological systems. For example, surfactant lining on the alveoli walls in the lungs increase the pulmonary or lung compliance thereby preventing the collapse of lungs after expiration. In case surfactant deficiency occurs, a replacement for the lung surfactant is used and its very important to understand efficiency of its spread and the stability of the newly created film for effectively treating the disorder [5]. Primitive micro-organisms such as bacteria are also know to exploit the same physical principle for facilitating the expansion of bacterial laden drops by lowering the surface tension to colonise various surfaces. Bacterial species such as *Bacillus subtilis* use this effect to climb walls of vessels containing bacteria-laden fluids [6] and *Pseudomonas aeruginosa* forms long straight fingers (also referred to as tendrils) as they collectively swarm on a nutrient rich agar surface as shown in supplementary figure s3 (a) which are attributed to the flow instabilities in surfactant driven films [7].

We are interested in understanding swarming in bacteria, which is a complex interplay of passive fluid dynamics and an "active" bacterial motility involving sensing of signalling molecules and consequent behavioural changes. Surfactant produced by the bacteria reduces the surface tension of water extracted from the agar and spreads into a thin water film in which the bacteria swarm. We performed experiments to study this phenomenon in the gram-negative bacteria *Pseudomonas aeruginosa* (PA).

Indeed, there have been several attempts to model the swarming pattern of PA based on a highly mechanistic multiscale model [8], as a population dispersal phenomenon using spatial kernels [9] and based on Marangoni forces [7,10,11]. The surfactant driven model proposed by Trinschek et al. [7] is able to explain several features of PA swarming pattern and its response to various perturbations.

The tendrils of a PA swarm colony seldom intersect and also avoid tendrils of the sister colony as shown in the supplementary figure s3 (b). The advancing tendrils also avoid inert physical obstacles made of Polydimethylsiloxane (PDMS) by going around the obstacle as shown in supplementary figure s3 (c). Although not reported previously, our simulations show that these behaviours can be described very well using the model proposed by Trinschek et al. [7] (see supplementary figure s3 (d,e,f)). This model considers the spreading surfactant laden film as a passive dispersal agent of the growing bacterial population and does not account for the ability of sensing and motility of the individual bacteria, and in that sense is an incomplete description of the phenomenon.

To understand the role of bacterial sensing and individual motility in swarming, experiments were designed with an intention to identify behaviours which could not be captured by the Trinschek et al. model [7] We reasoned that the best way to identify such behaviours will be employ perturbations which are not expected to interfere with the surfactant flow dynamics. To this end, we used antibiotic agents to perturb the swarming pattern.

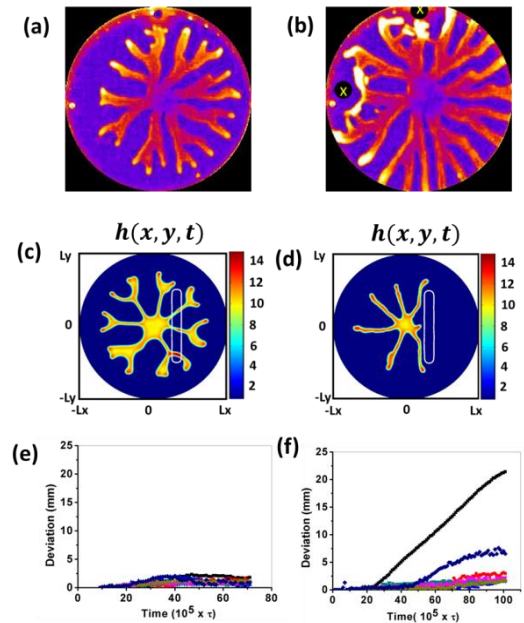

FIG1: Swarm colony of wild-type *Pseudomonas aeruginosa* (a) just before addition of antibiotic (b) after the addition of antibiotic ( yellow cross represents the positions at which antibiotic was added) , (c) simulated swarm colony using modified model (see supplementary section III) proposed by ref. 7 (d) simulated swarm colony using our proposed model at time $t = 7 * 10^6 \tau, [L_X, L_Y] = [4000, 4000]$ (the white contour in (c) and (d) represent the region where the antibiotic concentration is half of the maximum that has been added). The system was simulated using equations (1-3) in the presence and absence of the antibiotic and the deviation caused by it has been quantified for both the model in (e)(f).[$\tau$ is the time scaling ,Lx= x/L where L is length scaling parameter used in simulations and is described in supplementary section II] [Note: False color has been used to enhance the contrast of the images of agar plates from our experiments]

These experiments consisted of adding a $2\mu L$ drop of antibiotic gentamicin from 50 $mg/mL$ stock solution at the tip of one of the growing tendrils of a wild-type PA as shown in figure 1(a). We expected death of the bacteria at the tip and thereby leading to termination of the tip. But to our surprise, the bacterial tendrils changed the direction of growth and to the extent of intersecting with the neighbouring tendrils in direct contradiction to the behaviour seen in unperturbed swarm patterning (figure 1(b) and supplementary video Antibiotic_PA.avi). We used pure DI-water as control (figure not shown) in place of the antibiotic and observed no effect on the growing tendril thus ruling out the possibility of having induced any change in the physical properties of the agar or in surfactant concentration. This experiment provides a strong role of sensing and motility of the bacteria, in addition to surfactant gradient induced dispersal, in describing bacterial swarming behaviour. Thus, these two aspects, namely, surfactant driven fluid flow and active bacterial behaviour together determine the experimental observations, which are aptly described as motility driven surfactant assisted dispersal of bacteria.

We further investigated the swarming behaviour of flagella mutant (a strain without functional flagella and consequently immotile) on identical plates as those used for the wild-type PA. We found that although the flagella mutant produced similar concentrations of surfactants (Supported by LC-MS data in supplementary section VI(F) and supplementary figure s5), they were incapable of producing tendrils and instead the colony grew very slowly and symmetrically as shown in figure 2(b). Examination of single cell behaviour revealed active motility in the case of wild-type bacteria but essentially immotile situation in the case of flagella mutant, as expected. This indicates that the bacteria's ability to swim independently in the extracted water film is essential to create the observed swarming patterns. Absence of this ability results in an arrested swarm as in the case of the flagella mutant (See Supplementary videos Singlecell_PA.avi and Singlecell_flgM.avi).The materials and methods used in the experiments have been given in supplementary section VI.

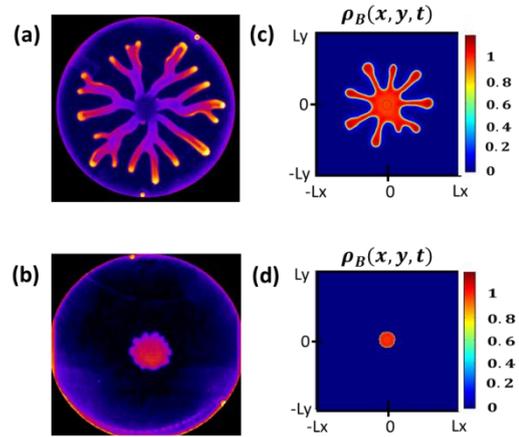

**FIG2.** swarm colony of (a) wild-type PA (b) flagella mutant PA and after 24 hours of inoculation. The system simulated using equations (1-3) indicate the bacterial density distribution for (c)$\chi = 1$ (d)$\chi = 0.01$ at $t = 5*10^6\tau$ ,$[L_X, L_Y] = [3000, 3000]$. [Note: False color has been used to enhance the contrast of the images of agar plates from our experiments]

It can be inferred from these experiments that the individual bacteria are continuously sensing their local environment and making decisions which affect their individual movements as well as the population distribution of the bacteria in the colony. The variation in population distribution can in turn modify the surface tension gradients in the film and affect the shape of the spreading film.

To quantitatively describe the interplay of the passive fluid physics and active bacterial behaviour we developed the following mathematical model. The mixture of the insoluble surfactant and bacteria in an incompressible thin film on the agar surface is modelled using the Navier-Stokes equation with the lubrication approximation because the $h(x,y,t)$ height of the water film is very small compared to the characteristic length L over which it varies. The concentration of the surfactant (area density) $\Gamma(x,y,t)$ is assumed to be below the critical micelle concentration and the surfactant's diffusion constant is taken to be D in the liquid film. The surface tension ($\gamma$) of the film depends on the surfactant concentration as $\gamma = \gamma_0 - \frac{k_B T}{a^2}\Gamma$, [7] where $\gamma_0$ is the surface tension of water, a is the typical surfactant length scale and $k_B T$ is the thermal energy. The bacterial (number density) density $\rho_B(x,y,t)$ extracts water from the agar layer underneath through osmosis and also produces surfactants. The bacterial density varies within the film due to bacterial growth, population dispersal by the spreading film and fluctuations in individual motility in response to sensory information such as the presence of an antibiotic.

Considering the partial wetting [7,12] and capillary effects (gravity neglected) for the pressure of the fluid, shear stress due to surface tension gradient at the air-water interface and no slip at the agar surface, the velocity profiles of the thin film is given by

$$U = \frac{1}{\mu}\nabla P\left(\frac{z^2}{2} - hz\right) + \frac{z}{\mu}\nabla\sigma \quad \text{and}$$

$$P - P_{atm} = -\gamma_o \nabla^2 h + A\left(\frac{1}{h^3} - \frac{h_a^3}{h^6}\right)$$

Where $= \left(\frac{\partial}{\partial x}, \frac{\partial}{\partial y}\right)$, U=(u,v) = velocities in x and y directions respectively, $\gamma_o$ is the surface tension of water, A is the Hamaker constant, $\mu$ is the viscosity of the bacteria laden fluid, $P_{atm}$ is the atmospheric pressure and $h_a$ is the height of the pre-existing adsorption layer.

Using the horizontal velocity derived above, we get the following evolution equations for $h(x,y,t), \Gamma(x,y,t)$ and $\rho_B(x,y,t)$ as follows

$$\frac{\partial h}{\partial t} = \nabla\cdot\left(\frac{h^3}{3\mu}\nabla\left(A\left(\frac{1}{h^3} - \frac{h_a^3}{h^6}\right) - \gamma_o\nabla^2 h\right)\right) + \frac{k_B T}{a^2}\nabla\cdot\left(\left(\frac{h^2}{2\mu}\right)\nabla\Gamma\right) + f(\rho_B, h) \quad (1)$$

$$\frac{\partial \Gamma}{\partial t} = \nabla\cdot\left(\frac{h^3}{3\mu}\nabla\left(A\left(\frac{1}{h^3} - \frac{h_a^3}{h^6}\right) - \gamma_o\nabla^2 h\right)\right) + \frac{k_B T}{a^2}\nabla\cdot\left(\left(\frac{h\Gamma}{\mu}\right)\nabla\Gamma\right) + \nabla\cdot(D\nabla\Gamma) + p(\rho_B, \Gamma) \quad (2)$$

$$\frac{\partial \rho_B}{\partial t} = \nabla\cdot\left(\frac{\chi\rho_B h^2}{3\mu}\nabla\left(A\left(\frac{1}{h^3} - \frac{h_a^3}{h^6}\right) - \gamma_o\nabla^2 h\right)\right) + \nabla\cdot\left(\frac{\chi h \rho_B}{2\mu}\frac{k_B T}{a^2}\nabla\Gamma\right) + \nabla\cdot(\eta\chi\rho_B\nabla c_a) + G(\rho_B) \quad (3)$$

The factor $\chi$ in the equation (3) represents the population fraction capable of active motility, described by their ability to swim independent of the surfactant dispersal forces, for instance as a response to a chemotactic signal. Based on this definition, the flagella mutant will be described with $\chi = 0$.

The individual bacteria sense their local environments for chemical cues (e.g. antibiotics) and modify their swimming directions. The average macroscopic flux resulting from the action of such chemical cues is modelled using the Keller-Segel form [13] i.e. $J = (\eta\chi\rho_B\nabla c)$ where $\eta$ is the chemotactic sensitivity of the bacteria, c is the concentration of chemical signalling molecule and $\chi\rho_b$ is the fraction of the motile bacteria.

The growth of the bacteria in equation (3) is modelled using a logistic equation

$$G(\rho_B) = g\rho_B(1-\rho_B)\Theta(\rho_B - \rho_{mn}) \quad (4)$$

Where g is the doubling rate of the bacteria, the step function Θ and the minimum bacterial density $\rho_{mn}$ ensures at least one bacterium is present which is necessary for the growth of population.

The osmosis of water by the bacteria in equation (1) is modelled as

$$f(\rho_B, h) = f_w (h_{mx}\rho_B - h)\Theta(h - h_a) \quad (5)$$

Where $\Theta(h - h_a)$ stabilises the adsorption layer of height $h_a$ in the absence of bacteria and $f_w$ is equal to the growth rate of the bacteria. Trinschek et al. [7] showed that the osmosis is a very fast process limited by the osmolyte production which is limited by the doubling rate g, of the bacteria. $h_{mx} \rho_B$ gives the equilibrium height of the film which is required for the given density of the bacteria in the location. $h_{mx}$ is the maximum height of the colony measured experimentally. The surfactant production by the bacteria in equation (2) is proportional to the bacteria density and is modelled as

$$p(\rho_B, \Gamma) = p_s\rho_B(\Gamma_{mx} - \Gamma)\Theta(\Gamma_{mx} - \Gamma) \quad (6)$$

Here $\Gamma_{mx}$ limits the surfactant production to match with the experimentally measured decrease in surface tension and $p_s$ is the production rate of the surfactant molecule. [6].

We use the non-dimensional form of the above equations for numerical simulations. The derivation of the evolution equation, non-dimensionalising factors, initial conditions, simulation details and the estimation of different parameters from the experiments have been provided in greater detail in the supplementary sections I , II and V.

The simulation results of our model figure 1(d) is able to capture the directional change in the presence of the antibiotic seen in the experiments. The original Trinschek et. al model [7], cannot account for the presence of the antibiotic and would produce a pattern resembling figure 1(c). To account for the effect of inhibition of bacterial growth due to the antibiotic, we modified the original growth rate term of their model as described in detail in the supplementary section III. However, even this modification did not result in swarming patterns resembling experimental observations. The quantified deviation in the direction of each tendril as given by figure 1 (e)(f) shows no drastic change in the pattern for the modified Trinschek et. al [7]model while there is a significant change in the pattern in the presence of antibiotic using our model which captures the experimental observations.

We then investigated whether the arrested pattern of the flagella mutant swarms can be reproduced by our model. Trinschek et al [7] also showed such arrested patterns in surfactant limited situations. However, in the case of PA, the flagella mutant as well as the wild-type produce equal concentrations of surfactants and therefore surfactant limitation cannot provide an explanation for the arrested pattern. In our model, we can reproduce the arrested pattern without requiring surfactant limited conditions by simply setting the motility parameter $\chi$ close to zero, as one would expect in the case of immotile flagella mutants (see figure 2(d), supplementary figure s1 and s2). The choice of $\chi = 0$ is also supported by the limited surface motility of these bacteria as seen in the single cell videos (See supplementary video Singlecell_flgM.avi)

In order to better capture the interplay of fluid physics, represented by surfactant concentration $\Gamma_{mx}$, and active bacterial behaviour, represented by the motility parameter $\chi$, we constructed a morphological phase diagram as shown in figure 3 (corresponding swarm patterns are shown in figure s2). The phase diagram is obtained using the value of circularity of the swarming pattern of obtained for various choices of $\Gamma_{mx}$ and $\chi$. The circularity for a pattern is given by

$$circularity = 4 * \pi * \frac{Area}{Perimeter^2} \quad (7)$$

The lower values of circularity indicate branched patterns and higher value of circularity indicates arrested patterns. The phase diagram shows that arrested patterns (with high circularity values) can be obtained either by surfactant limitation, as discussed by Trinschek et. al [7] or low motility as shown in our model and experiments with immotile bacteria.

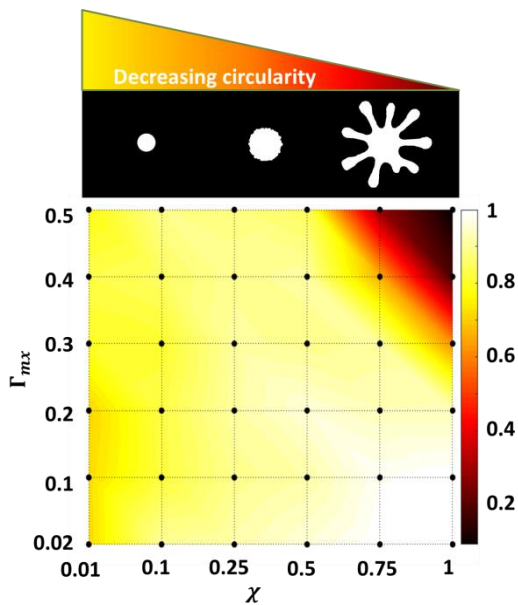

FIG3. $\Gamma_{mx}$ Vs $\chi$ phase diagram with grid points indicating the circularity of the simulated colony pattern obtained for the corresponding values of $\Gamma_{mx}$ and $\chi$. (Rest of the phase space is constructed by interpolation)

Finally, the form of coupling active motility and surfactant driven fluid dynamics as represented by equations (1) and (2) has a universal character. To illustrate this, we considered swarming behaviour in a different species, namely, *Bacillus subtilis*. In this species, it is known that the surfactant produced by *Bacillus subtilis* also acts as the osmotic agent [14]. We were able to reproduce the swarming pattern exhibited by *Bacillus subtilis* using the same set of equations (1),(2) and (3) describing activity-fluid dynamics coupling by using bacteria specific growth parameters as shown in figure 4 (also see supplementary figure s4). Our simulations are able to reproduce observations reported previously [15]. The generation of species specific parameters for the simulation is described in greater detail in the supplementary section IV.

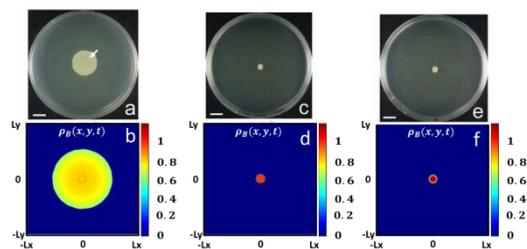

FIG4. Swarming pattern for a) wild-type (c) flagella mutant (e) Surfactant mutant of *Bacillus subtilis* have been reproduced from reference [15] . The simulated bacterial density $(\rho_B)$ was obtained using *B. subtilis* specific growth parameters in equations (1-3) for (b) wild-type (d) flagella mutant (f) surfactant mutant $at\ t = 1 * 10^6 \tau$, $[L_X, L_Y] = [3000, 3000]$.

.

In summary, motility of the bacteria plays an important role that has not been accounted for in the existing models of bacterial swarming. The ability of the bacteria to swim in the osmotically produced water film also enables their dispersal driven by surface tension gradients. The arrested swarm pattern of the flagella mutant suggests that in the absence of independent ability to swim, population dispersal is suppressed, thus reiterating the importance of individual motility. Our model accounts for the effect of active motility using the parameter $\chi$. Low $\chi$ may arise due to bacteria getting immobilised due to entrapment at the agar surface [16]. The negligible flow velocities at the boundary due to no slip at the agar surface reduce the passive dispersal of the bacteria that lie immobilised on the agar surface. Thus functional flagellum is necessary so that the bacteria can actively disperse themselves in the water film and facilitate long range passive dispersal. This idea may also be used to arrest expansion of swarming in bacteria by creating surfaces that deters its motility by entrapping it. Overall, our model provides a more complete picture of bacterial swarming including the effect of bacterial sensing and active motility in response to sensory information.

We gratefully acknowledge Robert Bosch Centre for Cyber Physical Systems at Indian institute of Science, Bangalore, India for funding this research. We also acknowledge the use of facilities at Centre for Nano Science and Engineering, Indian Institute of Science, Bangalore, India.

# Supplementary Material: Active modulation of surfactant driven flow instabilities by swarming bacteria

Harshitha S. Kotian[1][2], Shalini Harkar[2], Shubham Joge[3], Ayushi Mishra[3], Amith Z. Abdulla[1], Varsha Singh[3], Manoj M. Varma[1][2]*

[1] Centre for Nano Science and Engineering, Indian Institute of Science, Bangalore, India
[2] Robert Bosch Centre for Cyber Physical Systems, Indian Institute of Science, Bangalore, India
[3] Molecular Reproduction, Development and Genetics, Indian Institute of Science, Bangalore, India
*mvarma@iisc.ac.in

### I. Derivation of evolution equations of our model

Consider a thin adsorption layer of water of height $h_a$ on nutrient rich agar inoculated with bacteria in a small area at the centre. These bacteria possibly produce osmolytes that help them form a thin water film on the surface of agar in which they can swim. The height of this water film created is h(x,y,t) and the surface of the film is covered by insoluble surfactant molecules (area density) $\Gamma(x, y, t)$ that the bacteria produces. The horizontal driving potential of this film has contribution from wetting energy [1,2] and capillary effect ( effect of gravity can be neglected) as follows

$$P - P_{atm} = -\gamma \nabla^2 h + A\left(\frac{1}{h^3} - \frac{h_a^3}{h^6}\right) \quad \text{(s1)}$$

Where γ is the surface tension, A is the Hamaker constant and $h_a$ is the height of the adsorption layer.

The velocity of the thin water film being slow over a long characteristic length gives the following reduced momentum equations

$$-\frac{\partial P}{\partial x} + \frac{\mu \partial^2 u}{\partial z^2} = 0$$

$$-\frac{\partial P}{\partial y} + \frac{\mu \partial^2 v}{\partial z^2} = 0$$

$$-\frac{\partial P}{\partial z} = 0 \quad \text{(s2)}$$

Where u,v are the velocity of the film in x and y direction respectively and $\mu$ is the viscosity of bacterial colony.

Considering the tangential stress at the liquid-air boundary due to gradient in surface tension and no slip condition at liquid-solid boundary, we get the following boundary conditions.

Assume $U = u\hat{i} + v\hat{j} + w\hat{k}$

$$\frac{\mu \partial u}{\partial z}\bigg|_{z=h} = \frac{\partial \gamma}{\partial x}$$

$$\frac{\mu \partial v}{\partial z}\bigg|_{z=h} = \frac{\partial \gamma}{\partial y}$$

$$u, v|_{z=0} = 0 \qquad (s3)$$

Where $\gamma$ is the net surface tension resulting from the presence of insoluble surfactant molecules.

Solving for (u,v) using the reduced momentum equations (s2) with the boundary conditions (S3) mentioned above, we get

$$(u, v) = \frac{1}{\mu} \nabla P \left(\frac{z^2}{2} - hz\right) + \frac{z}{\mu} \nabla \gamma \qquad (s4)$$

Where $\nabla = (\frac{\partial}{\partial x}, \frac{\partial}{\partial y})$

Using the incompressibility of the flow, we get

$$\frac{\partial h}{\partial t} + \left(\int_0^h u \, dz\right)_x + \left(\int_0^h v \, dz\right)_y = 0 \qquad (s5)$$

Substituting for velocity (equation s4) and driving potential (equation s1), we obtain equation for height of the film h(x,y,t) as

$$\frac{\partial h}{\partial t} = \nabla \cdot \left(\frac{h^3}{3\mu} \nabla \left(A\left(\frac{1}{h^3} - \frac{h_a^3}{h^6}\right) - \gamma \nabla^2 h\right)\right) - \nabla \cdot \left(\left(\frac{h^2}{2\mu}\right) \nabla \gamma\right) \qquad (s6)$$

And the spatial distribution of the surfactant molecules is given by

$$\frac{\partial \Gamma}{\partial t} + \nabla \cdot (U\Gamma) = \nabla \cdot (D \nabla \Gamma) \qquad (s7)$$

Where D is the diffusion constant and $U = (u, v)$

To fully couple the height of the water film and surfactant concentration, we relate the net surface tension to the surfactant concentration [1] as given below.

$$\gamma = \gamma_0 - \left(\frac{k_B T}{a^2} \Gamma\right) \qquad (s8)$$

$$\frac{\partial \gamma}{\partial x} = -\left(\frac{k_B T}{a^2} \frac{\partial \Gamma}{\partial x}\right) \qquad (s9)$$

Where $k_B T$ is the thermal energy and $a^2$ is the effective area occupied by the surfactant molecule at the interface. The effect of surface tension gradients can be neglected on capillarity i.e. $\gamma \approx \gamma_o$.

The overall equations by substituting for velocity and surface tension (equations (s4) and (s9)) in equations for height $h(x, y, t)$ and surfactant concentration $\Gamma(x, y, t)$ (equations (s6) and (s7))

$$\frac{\partial h}{\partial t} = \nabla \cdot \left(\frac{h^3}{3\mu} \nabla \left(A\left(\frac{1}{h^3} - \frac{h_a^3}{h^6}\right) - \gamma_o \nabla^2 h\right)\right) + \frac{k_B T}{a^2} \nabla \cdot \left(\left(\frac{h^2}{2\mu}\right) \nabla \Gamma\right)$$
$$(s10)$$

$$\frac{\partial \Gamma}{\partial t} = \nabla \cdot \left( \frac{h^2 \Gamma}{3\mu} \nabla \left( A \left( \frac{1}{h^3} - \frac{h_a^3}{h^6} \right) - \gamma_o \nabla^2 h \right) \right) + \frac{k_B T}{a^2} \nabla \cdot \left( \left( \frac{h\Gamma}{\mu} \right) \nabla \Gamma \right) + \nabla \cdot (D \nabla \Gamma)$$

(s11)

These equations (s10 and s11) are purely hydrodynamic equations and hence the passive part of the system.

The bacteria along with random motion also get driven by movement of the water film which in turn is driven by surfactant gradients. The active behaviour of the bacteria is defined by the term motility parameter $\chi$. The bacteria is represented by $\rho_B(x,y,t)$ (number density) in a given position normalised with the maximum that can be sustained by the nutrients in the agar.

$$\frac{\partial \rho_B}{\partial t} + \nabla \cdot (\chi U_{avg} \rho_B) = \nabla \cdot (\chi \eta \rho_B \nabla c) + G(\rho_B) \qquad (s12)$$

Where $\eta$ the sensitivity of the bacteria to antibiotic and c is the antibiotic concentration

Since the bacteria swims in the water film created, the drift velocity can be obtained by averaging the horizontal velocity (equation s4) across the h(x,y,t). This average velocity is given by

$$U_{avg} = \left( \frac{-h^2}{3\mu} \nabla \left( A \left( \frac{1}{h^3} - \frac{h_a^3}{h^6} \right) - \gamma_o \nabla^2 h \right) \right) - \frac{h}{2\mu} \frac{k_B T}{a^2} \nabla \Gamma \qquad (s13)$$

The active part of the system can be divided into three subsystems: growth rate $G(\rho_B)$ and motility of the bacteria ($\chi$), Osmosis of water $f(\rho_B, h)$ by bacteria produced osmolytes, production of surfactant $p(\rho_B, \Gamma)$. The mathematical forms of these functions have been mentioned in the main text.

Now, combining the active and the passive components of the system, we obtain the following equations

$$\frac{\partial h}{\partial t} = \nabla \cdot \left( \frac{h^3}{3\mu} \nabla \left( A \left( \frac{1}{h^3} - \frac{h_a^3}{h^6} \right) - \gamma_o \nabla^2 h \right) \right) + \frac{k_B T}{a^2} \nabla \cdot \left( \left( \frac{h^2}{2\mu} \right) \nabla \Gamma \right) + f(\rho_B, h) \qquad (s14)$$

$$\frac{\partial \Gamma}{\partial t} = \nabla \cdot \left( \frac{h^3}{3\mu} \nabla \left( A \left( \frac{1}{h^3} - \frac{h_a^3}{h^6} \right) - \gamma_o \nabla^2 h \right) \right) + \frac{k_B T}{a^2} \nabla \cdot \left( \left( \frac{h\Gamma}{\mu} \right) \nabla \Gamma \right) + \nabla \cdot (D \nabla \Gamma) + p(\rho_B, \Gamma) \qquad (s15)$$

$$\frac{\partial \rho_B}{\partial t} = \nabla \cdot \left( \frac{\chi \rho_B h^2}{3\mu} \nabla \left( A \left( \frac{1}{h^3} - \frac{h_a^3}{h^6} \right) - \gamma_o \nabla^2 h \right) \right) + \nabla \cdot \left( \frac{\chi h \rho_B}{2\mu} \frac{k_B T}{a^2} \nabla \Gamma \right) + \nabla \cdot (\chi \eta \rho_B \nabla c) + G(\rho_B) \qquad (s16)$$

$$\frac{\partial c}{\partial t} = \nabla \cdot (D_a \nabla c) \qquad (s17)$$

## II. Non dimensionalisation and experimentally determined parameter value

| Symbol | Meaning | Experimental Value |
|---|---|---|
| $\mu$ | Viscosity of the bacterial laden film | $0.1\ Pa\ s$ [ref. [3]] |
| $h_a$ | Height of pre-existing adsorption layer | $1\mu m$ [ref. [1]] |
| $a$ | Surfactant length scale | $3nm$ [ref. [1]] |
| $k_B T$ | Energy at $37^oC$ | approx. $4*10^{-21} J$ |
| $\gamma_o$ | Surface tension of water at $37^oC$ | $70\ mN\ m^{-1}$ |
| G | Doubling time of *P. aeruginosa* in minimal media | 1-1.5 hours [ref. [4]] |
| $h$ | Average colony height of *P.aeruginosa* | $10\mu m$ [ref. [3]] |
| D | Translational diffusion constant of the surfactant | $3*10^{-10} m^2\ s^{-1}$ <br> $[1.8 - 6.7 * 10^{-10}\ m^2 s^{-1}$ ref. [5]] |
| $\dot\gamma$ | Decrease in surface tension due to surfactant production | $-0.18\ mN/(m.hr)$ <br> [ref. [3]] |
| $D_B$ | Diffusion coefficient of motile cells | $1.5 * 10^{-9} m^2 s^{-1}$ <br> [ref. [6]] |
| $\eta$ | Chemotactic sensitivity coefficient | $1.5 - 75 * 10^{-9} m^2 s^{-1}$ <br> [ref. [6]] |

To obtain a dimensionless form of the model, we use the following scaling as suggested by [1]

| Scaling and scaled parameters | Estimated value |
|---|---|
| Surface tension scale ($\kappa$) | $\kappa = \dfrac{k_B T}{a^2} = 4*10^{-4}\ N/m$ |

| Vertical length scale ($l$) | $l = h_a$ |
| --- | --- |
| Horizontal length scale ($L$) | $L = \sqrt{\frac{\gamma_o}{\kappa}} l = 10 \mu m$ |
| Time scale ($\tau$) | $\tau = \frac{L^2 \mu}{\kappa l} = 0.03 \, s$ |
| Dimensionless doubling or growth rate ($\tilde{g}$) | $\tilde{g} = g\tau = 0.833 * 10^{-5} \sim 10^{-5}$ |
| Dimensionless colony height ($\tilde{h}$) | $\tilde{h} = \frac{h}{l} = 10$ |
| Dimensionless surfactant production rate ($\widetilde{p_s}$) [$\Gamma_{mx}$ = maximum surfactant concentration] | $\dot{\Gamma} = -\frac{\dot{\gamma}\tau}{\kappa} = 0.375 * 10^{-5}$ $\dot{\Gamma} = \widetilde{p_s} \Gamma_{mx}$ Where $\widetilde{p_s} = 10^{-5}$ and $\Gamma_{mx} = 0.5$ and $0.02$ for Wild-type PA and surfactant mutant respectively [ref. [1]] |
| Wettability parameter(W) (Note: A=Hamaker constant) | $W = \frac{A}{\kappa h_a^2}$ , W=0.05 [ref. [1]] |

The antibiotic experiments involved introducing $2 \mu l$ of $50 \, mg/ml$ stock solution of gentamicin such that the total amount at the point of introduction on the agar plate is $100 \, \mu g$. MIC (minimum inhibition concentration) for gentamicin is in the range $0.03 - 64 \, \mu g/ml$ [7] which translates to $0.75 - 1600 \, \mu g$ of gentamicin distributed throughout the agar (25 ml of agar used per plate). The deviation in the direction of tendril propagation in the presence of diffusing antibiotic has been observed in our experiments. This indicates that the bacteria are alive, growing and the concentration at these distances is less than the MIC concentration. The dimensionless initial concentration of antibiotic is assumed to be 10 and the threshold concentration for the growth inhibition ($A_{thresh}$) is taken as 1. A low value of the diffusion constant of antibiotic is taken ($\widetilde{D_a}$=0.001) in simulations to ensure that the antibiotic does not spread appreciably during the simulation period. This match with the experimental conditions where the antibiotic is added after swarming has initiated and the tendril patterns are fully developed. Therefore the diffusion length of the antibiotic is low compared to that of the surfactant molecules although they would be expected to have the same diffusion constant based on molecular sizes.

As seen in our experiments, tendril width does not change significantly within the duration of experiment indicating that the diffusion or zero gradient random walk of the bacteria must be confined within the colony. The effect of cell diffusion on the pattern was negligible (supplementary figure s6) and is assumed zero for the simulations. The chemotactic sensitivity to the antibiotic measures the change in dispersal capabilities of the bacterial population for unit change in concentration of the antibiotic. The value of chemotactic sensitivity is taken as $\tilde{\eta} = 0.01$ ($\eta = 3 * 10^{-11} m^2/s$) in our simulations which is less than the experimental value as movement of bacteria is restricted by the height of the water film.

Bacterial density at the tips has been measured to be $1.5 - 2 * 10^6/mm^2$. Assuming the area occupied by the bacteria $1 \mu m^2$, we get 1-2 bacteria/$\mu m^2$ distributed along the height of the film. Thus, the minimum bacteria density is taken as $\rho_{mn} = 0.5$ (maximum bacterial density $\rho_{mx} = 1$) so that at least one bacterium is present which is necessary for the growth of population.

The velocity of the propagating tendril tip has been experimentally found to be ~50 $\mu m/min$ and the velocity of tendril tip obtained through simulations is ~6$\mu m/min$ which is about an order of magnitude less and this may be due to overestimation of viscosity and unaccounted spatial variations in viscosity which depend on cell densities [8].

**Non-dimensional form of our model**

$$\frac{\partial \tilde{h}}{\partial \tau} = \nabla. \left( \frac{\tilde{h}^3}{3} \left( \tilde{\nabla} \left( W \left( \frac{1}{\tilde{h}^3} - \frac{1}{\tilde{h}^6} \right) - \tilde{\nabla}^2 \tilde{h} \right) \right) + \tilde{\nabla}. \left( \left( \frac{\tilde{h}^2}{2} \right) \tilde{\nabla} \Gamma \right) + \tilde{f}(\tilde{h}, \rho_B)$$

$$\frac{\partial \Gamma}{\partial \tau} = \tilde{\nabla}. \left( \frac{\Gamma \tilde{h}^2}{3} \tilde{\nabla} \left( W \left( \frac{1}{\tilde{h}^3} - \frac{1}{\tilde{h}^6} \right) - \tilde{\nabla}^2 \tilde{h} \right) \right) + \tilde{\nabla}. \left( (\tilde{h} \Gamma) \tilde{\nabla} \Gamma \right) + \tilde{\nabla}. (\tilde{D} \tilde{\nabla} \Gamma) + \tilde{p}(\rho_B, \Gamma)$$

$$\frac{\partial \rho_B}{\partial \tau} = \nabla. \left( \frac{\chi \rho_B \tilde{h}^2}{3} (\tilde{\nabla} \left( W \left( \frac{1}{\tilde{h}^3} - \frac{1}{\tilde{h}^6} \right) - \tilde{\nabla}^2 \tilde{h} \right) \right) + \tilde{\nabla}. \left( \left( \frac{\chi \rho_B \tilde{h}}{2} \right) \tilde{\nabla} \Gamma \right) + \nabla. (\chi \tilde{\eta} \rho_B \nabla c) + \tilde{G}(\rho_B)$$

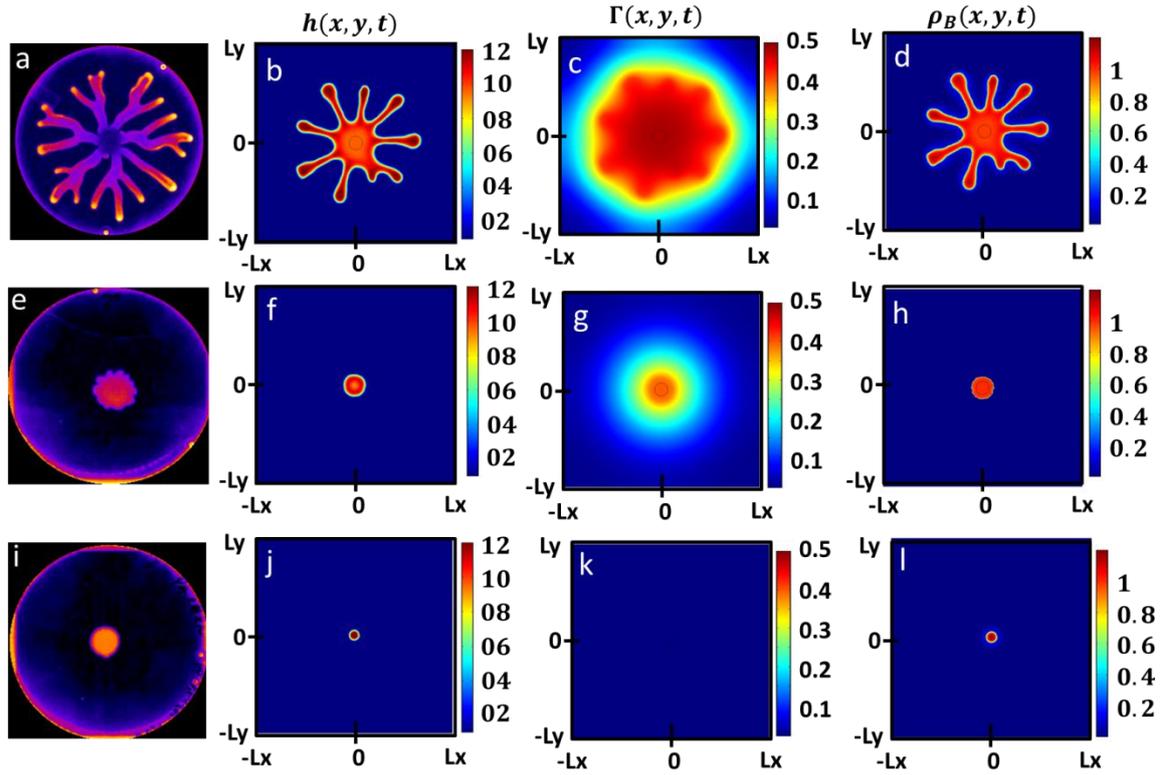

FIG s1.(a) Wild type PA colony (c) Flagella mutant(flgM) PA colony (e) surfactant mutant(rhlR) PA colony cultured on PGM-0.6% agar plates over 24 hours. The system has been simulated for height h(x,y,t), surfactant concentration $\Gamma(x,y,t)$ and bacterial density $\rho_B(x,y,t)$ using equations (1-3) with parameters (b,c,d) $[\Gamma_{mx}=0.5, \chi=1]$; (f,g,h) $[\Gamma_{mx}=0.5, \chi=0.01]$ ; (j,k,l)$[\Gamma_{mx}=0.02, \chi=1]$ for wildtype PA, Flagella mutant and surfactant mutants respectively.$[t=5*10^6\tau, [L_x, L_y]=[3000,3000], \rho_{mn}=0.5]$ [Note: False color has been used to enhance the contrast of the images of agar plates from our experiments]

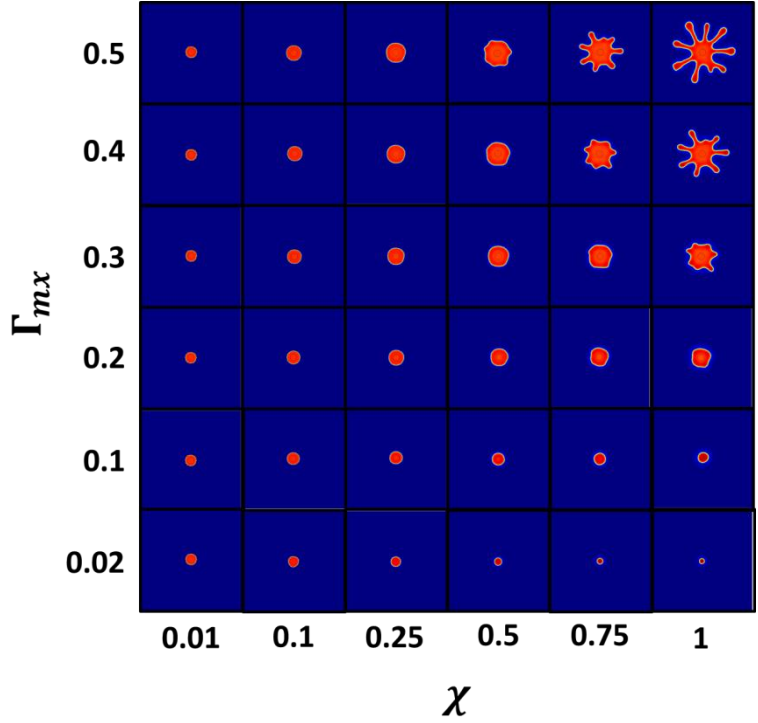

**FIG s2.:** Swarming patterns $\rho_B(x, y, t)$ obtained for various $\Gamma_{mx}$ and $\chi$ by simulating the system described by equations (1-3) in the main text.

### III. Modification to the model proposed by Trinschek et al [1].

Evolution equations:

$$\frac{\partial h}{\partial t} = \nabla \cdot \left( \frac{h^3}{3\mu} \nabla \left( A \left( \frac{1}{h^3} - \frac{h_a^3}{h^6} \right) - \gamma_o \nabla^2 h \right) \right) + \frac{k_B T}{a^2} \nabla \cdot \left( \left( \frac{h^2}{2\mu} \right) \nabla \Gamma \right) + G(h)$$

$$\frac{\partial \Gamma}{\partial t} = \nabla \cdot \left( \frac{h^3}{3\mu} \nabla \left( A \left( \frac{1}{h^3} - \frac{h_a^3}{h^6} \right) - \gamma_o \nabla^2 h \right) \right) + \frac{k_B T}{a^2} \nabla \cdot \left( \left( \frac{h\Gamma}{\mu} \right) \nabla \Gamma \right) + \nabla \cdot (D \nabla \Gamma) + P(h, \Gamma)$$

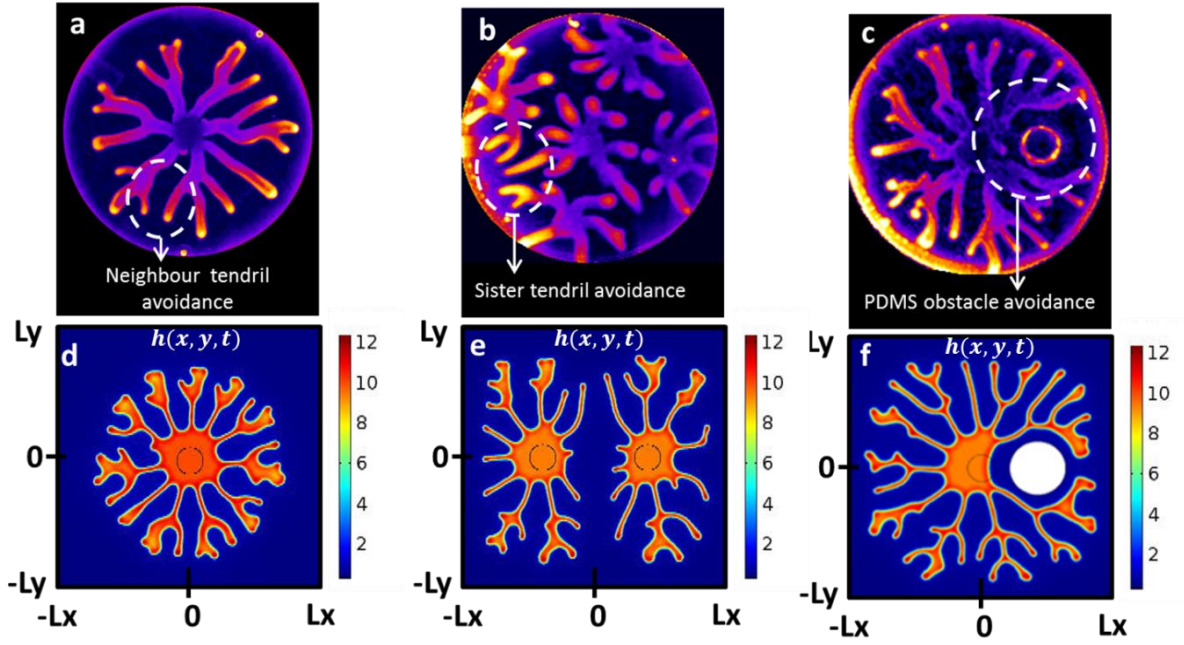

FIG s3: .(a) Wild type PA colony (b) Multiple wild type PA colonies (c) wild type PA colony with PDMS obstacle on PGM-0.6% agar plates over 24 hours. The experiment equivalent conditions have been simulated using model given by reference [1] where obstacle is given by a no-flux region with in the simulating arena. We find close agreement of experimental with simulation data [W=0.05, $t = 7.5 * 10^6 \tau, [L_X, L_y] = [5000, 5000]$] [Note: False color has been used to enhance the contrast of the images of agar plates from our experiments]

Modified G(h) and P(h, Γ):

$$G(h,c) = gh\left(1 - \frac{h}{h_{mx}}\right) \Theta'(h - h_{mn}) \Theta'(A_{thresh} - c)$$

$$P(h, \Gamma, c) = ph(\Gamma_{mx} - \Gamma) \Theta(\Gamma_{mx} - \Gamma) \Theta(h - h_{mn}) \Theta(A_{thresh} - c) \quad \text{(s18)}$$

where function $\Theta'(x) = \begin{cases} -0.05, x < 0 \\ 1, x \geq 0 \end{cases}$

The above growth rate of height of the film and production of surfactant have been modified to account for growth inhibition of the bacteria in the region where concentration c of antibiotic is greater than threshold concentration $A_{thresh}$. $h_{mn}(5\mu m)$ is the minimum height of the film that contains atleast one bacterium. $\Gamma_{mx}$ is the maximum surfactant concentration. Note that the $\Theta'$ is the the modified step function to prevent the colony proliferation in the absence of bacteria.

IV. **Model tuned for *Bacillus substilis***

$$h_t = \nabla \cdot \left(\frac{h^3}{3\mu} \nabla \left(A\left(\frac{1}{h^3} - \frac{h_a^3}{h^6}\right) - \gamma_o \nabla^2 h\right)\right) + \frac{k_B T}{a^2} \nabla \cdot \left(\left(\frac{h^2}{2\mu}\right) \nabla \Gamma\right) + f(\rho_B, h)$$

$$\frac{\partial \Gamma}{\partial t} = \nabla \cdot \left( \frac{h^3}{3\mu} \nabla \left( A \left( \frac{1}{h^3} - \frac{h_a^3}{h^6} \right) - \gamma_o \nabla^2 h \right) \right) + \frac{k_B T}{a^2} \nabla \cdot \left( \left( \frac{h\Gamma}{\mu} \right) \nabla \Gamma \right) + \nabla \cdot (D \nabla \Gamma) + p(\rho_B, \Gamma)$$

$$\frac{\partial \rho_B}{\partial t} = \nabla \cdot \left( \frac{\chi \rho_B h^2}{3\mu} \nabla \left( A \left( \frac{1}{h^3} - \frac{h_a^3}{h^6} \right) - \gamma_o \nabla^2 h \right) \right) + \nabla \cdot \left( \frac{\chi h \rho_B}{2\mu} \frac{k_B T}{a^2} \nabla \Gamma \right) + G(\rho_B)$$

Tuning growth functions for the Bacillus specific parameters,

$$f(\rho_B, h) = f_w \left( h_{surf} \Gamma + h_{mx} \rho_B - h \right) \Theta(h - h_a) \qquad \text{(s19)}$$

where $h_{surf}$ is the equilibrium height of the water film produced due to strong osmolytic effect of the surfactant produced by Bacillus [9] [10] in addition to other osmolytes that it may produce which accounts for equilibrium height $h_{mx}$. The step function $\Theta(h - h_a)$ stabilised the adsorbing layer in the absence of bacteria or surfactant. For simulations, $h_{surf} = 100, h_{mx} = 10$ as the height of the *Bacillus subtilis* colony measured was 60-100 $\mu m$ [9] and rest of the parameters are taken similar to that of PA.

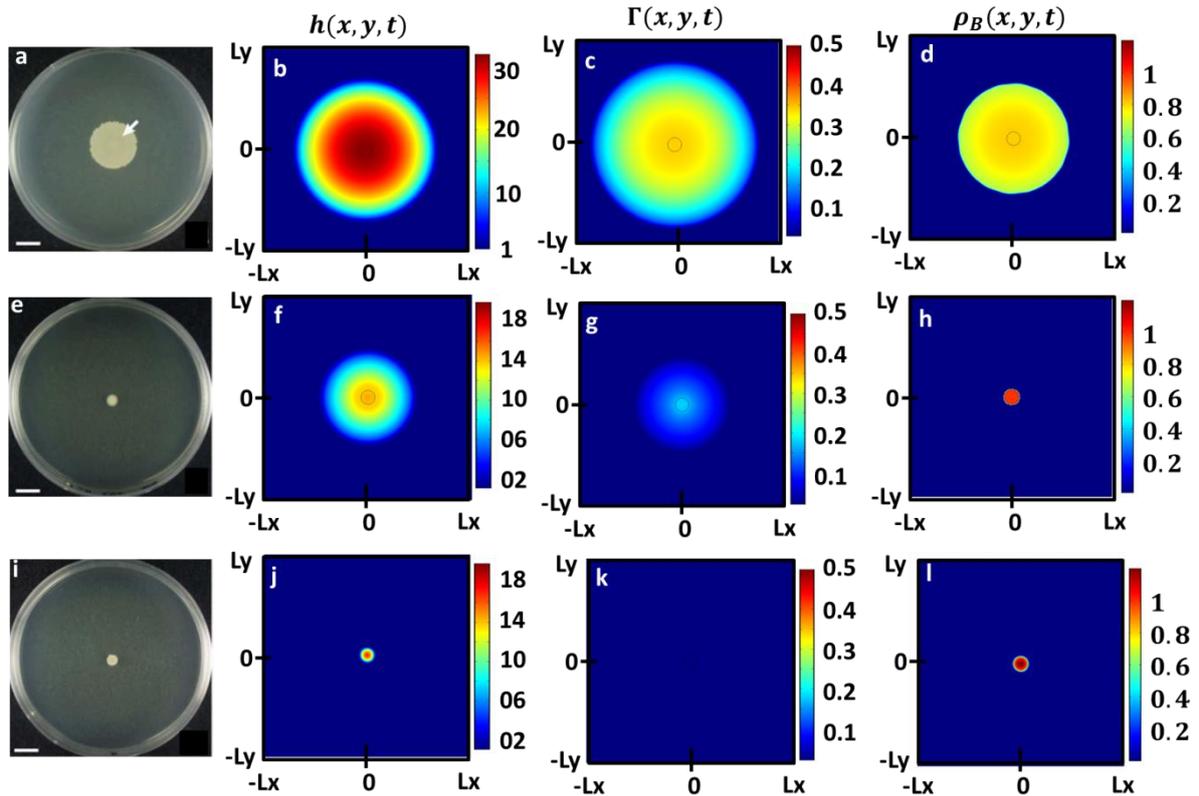

**FIG s4.** Experimental data for *B.Subtilis* has been taken from reference [9] and are presented here only for convenient comparison with the simulated pattern. (a) Wild type *B. Subtilis* colony (F29-3) (c) Flagella mutant (FW463) (e) Surfactant mutant (FK 955) cultured on LB-0.4% agar plates over 24 hours. The system has been simulated for height h(x,y,t), surfactant concentration $\Gamma(x, y, t)$ and bacterial density $\rho_B(x, y, t)$ using equations (1-3) using parameters (b,c,d) [$\Gamma_{mx} = 0.5, \chi = 1$]; (f,g,h) [$\Gamma_{mx} = 0.5, \chi = 0.01$]; (j,k,l)[$\Gamma_{mx} = 0.02, \chi = 1$] for wildtype PA, Flagella mutant and surfactant mutants respectively.[ $t = 1 * 10^6 \tau, [L_x, L_y] = [3000, 3000], \rho_{mn} = 0.2$].

## V. Simulation

Coefficient form PDE module of COMSOL multiphysics 5.0 was used for simulations. A free triangular extra fine mesh was calibrated for fluid dynamics. The simulation domain was defined by square $[-L_x, L_x] * [-L_y, L_y], (L_x, L_y) = [3000, 3000]$. Non-linear fully coupled Newton-Rhaphson method was used to solve the set of equations. The initial conditions of height was set to $h_a$ (height of adsorption water film) and surfactant concentration was set to 0 in the whole domain. At the beginning of the simulation, bacteria density was set to 1 within a circular region(radius=300) concentric with the simulation domain. The irregularity in the free triangular mesh provides the disturbance on the front that leads to fingered patterns that we see in the simulation.

## VI. Materials and Methods
### A. Swarming Motility Assay

For swarming assay, we used Peptone growth medium (PGM). Composition of PGM 0.6% agar plates are 6 grams of bacteriological agar (Bacto agar), 3.2 grams of peptone, and 3 grams of sodium chloride (NaCl) added in 1 litre of distilled water. The medium was autoclaved at 121°C for 30 minutes. After autoclaving the media, 1 mL of 1M CaCl2 (Calcium chloride), 1 mL of 1 M MgSO4 (Magnesium sulphate), 25 ml of 1M KPO4 and 1 mL of 5 mg/mL cholesterol were added into the medium and mixed properly. 25 mL PGM were poured in each 90mm Petri-plates and allowed them to solidify at room temperature (RT) for a half an hour under the laminar hood flow with the lid opened. And all the plates were kept at room temperature for 16-18 hours for further drying.

### B. Bacterial growth:
Bacteria are stored at -80 ˚C in 50 % glycerol stocks and antibiotic Carbenicillin was added to the frozen stocks of the bacteria. The cells were grown in LB Luria media with composition 10 g/L tryptone, 5 g/L yeast extract and 0.5 g/L sodium chloride, at 37 ˚C with continuous shaking for about 12 hours .
(Note that 12 hours growth in 3ml LB - Luria media at 37 ˚C would essentially yield a bacterial culture of optical density ranging from 2 to 3).

### C. Swarming

2ul of a planktonic culture of *Pseudomonas aeruginosa* strain PA14 with OD >2.8 is inoculated at the centre of 90 mm petri-plate containing PGM-0.6% agar. The wild type PA14 forms the pattern as shown in FIG. s1 a) over a period of 24 hours in a 90 mm petri-plate. *Pseudomonas aeruginosa* produces surfactants called Rhanmnolipids which are composed of monorhamnolipids (mono- RLs) , dirhamnolipids (di-RLs), and 3-(3- hydroxyalkanoyloxy) alkanoic acids (HAAs) [11]. The surfactant mutants are also referred to as Rhamnolipid mutants. All mutants (flagella mutant - flgM and rhamnolipid mutant- rhlR) are a transposon insertion mutants ,part of *P. aeruginosa* transposon insertion library [12].The inoculated plates were placed in an incubator at 37 ˚C. The growth of bacterial swarm colony was monitored for duration of about 24hours.

### D. Antibiotic Response

2µL of 50 mg/mL antibiotic Gentamicin was added to specific locations on the agar plate. This was done to observe the response of bacterial swarm to the localized presence of antibiotics in the medium. Addition of antibiotic was done after about 16 hours from the time of bacterial inoculation.

### E. Preparation of PDMS (Poly DiMethyl Siloxane) obstacle

We have used Sylgard 184 from Dow Corning. It has two parts: an elastomer part and the curing agent. The two parts i.e. elastomer: curing agent is mixed in the ratio of 10:1. This mixture is stirred well. The air bubble cause due to stirring is removed by degassing the PDMS mixture in a desiccator connected to vacuum pump. An acrylic template is made to obtain different shapes of the obstacle. The air bubble free PDMS mixture is then poured into the template and cured for 12 hours. The cured PDMS solidifies and is removed from the acrylic template. These PDMS obstacles are then sterilised in autoclave.

The sterilised obstacle blocks are placed in an appropriate position in the petri-plate. The nutrient agar is then poured around the obstacle such that the obstacle is half immersed in the nutrient agar while held intact in its original position. The nutrient agar with the obstacle is allowed to dry under the laminar hood.

### F. Liquid Chromatography- Mass Spectrometry

LC-MS study was done to analyse the amount of surfactant produced by different mutants (flgM and rhlR) and wild type. A reference standard sample of rhamnolipid (Mono-rhamnolipid dominant) was obtained from Sigma. The secretion by the bacterial samples was compared to the standard in order to quantify the amount of rhamnolipid present in different bacterial cultures. It was found that the flgM mutant and the wild-type had similar amounts of surfactant production while the rhlR mutant lacked rhamnolipid production. We have obtained characteristic peaks of di-rhamnolipid and mono-rhamnolipid [13] in the spectrometry.

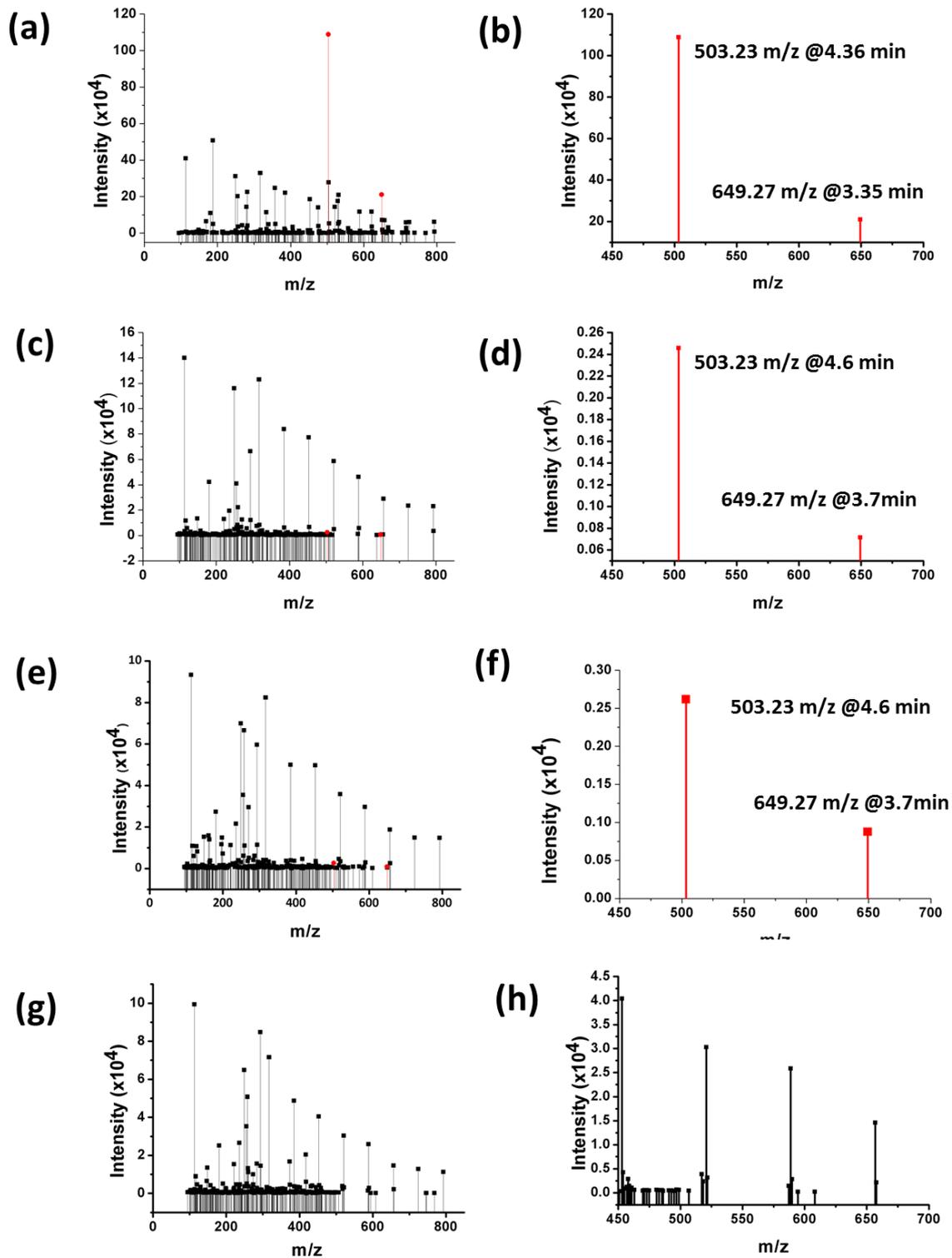

FIG s5: LC-MS data (a) for standard (mono-rhamnolipid dominant) (b) Intensity peaks of mono-rhamnolipid (503.23 m/z and retention time 4.36 min) and di-rhamnolipid (649.27 and retention time 3.35 min) ;LC-MS data and the corresponding intensity peaks for mono and di-rhamnolipid for (c)(d) wild type (e)(f)Flagella (flgM) mutant and (g)(h) Rhamnolipid (rhlR) mutant.